\documentstyle[aps,prl]{revtex} 
\draft 
 
\def\lsim{\mathrel{\raise.3ex\hbox{$<$\kern-.75em\lower1ex\hbox{$\sim$}}}} 
\def\gsim{\mathrel{\raise.3ex\hbox{$>$\kern-.75em\lower1ex\hbox{$\sim$}}}}

\begin{document} 
 
\twocolumn[\hsize\textwidth\columnwidth\hsize\csname
@twocolumnfalse\endcsname
 
\title {Can Supersymmetry Naturally Explain the Positron Excess?} 
\author{Dan Hooper$^1$, James E. Taylor$^1$ and Joseph Silk$^{1,2}$} 
\address{
$^1$ Astrophysics Department, University of Oxford, OX1 3RH  Oxford, UK;
$^2$ Institut d'Astrophysique de Paris, France}
\date{\today} 
 
\maketitle 
 
\begin{abstract}

It has often been suggested that the cosmic positron excess observed by the HEAT experiment could be the consequence of supersymmetric dark matter annihilating in the galactic halo. Although it is well known that evenly distributed dark matter cannot account for the observed excess, if substantial amounts of local dark matter substructure are present, the positron flux would be enhanced, perhaps to the observed magnitude. In this paper, we attempt to identify the nature of the substructure required to match the HEAT data, including the location, size and density of any local dark matter clump(s). Additionally, we attempt to assess the probability of such substructure being present. We find that if the current density of neutralino dark matter is the result of thermal production, very unlikely ($\sim 10^{-4}$ or less) conditions must be present in local substructure to account for the observed excess.

\end{abstract}

]

\section{Introduction}

A large body of evidence has accumulated in the favor of cold
dark matter. This body of evidence includes observations of galactic
clusters and large scale structure \cite{structure}, supernovae
\cite{supernovae} and the cosmic microwave background (CMB) anisotropies
\cite{cmb,wmap}. At the 2$\sigma$ confidence level, the
density of non-baryonic \cite{nonbaryonic}, and cold, dark matter is now known to be
$\Omega_{\rm{CDM}} h^2 = 0.113^{+0.016}_{-0.018}$ \cite{wmap}. 

A compelling dark matter candidate is provided
by supersymmetry \cite{susylsp}. In supersymmetric models which
conserve R-parity \cite{rparity}, the lightest supersymmetric particle
(LSP), is stable. Furthermore, in many supersymmetric models, the LSP
is the lightest neutralino, a mixture of the superpartners of the
photon, $Z$ and neutral Higgs bosons, and is electrically
neutral, colorless and, therefore, a viable dark matter candidate.   
If such a particle were in equilibrium with photons in the early universe,
as the temperature decreased, a freeze-out would occur leaving a thermal
relic density. The temperature at which this occurs, and
the density which remains, depends on the annihilation cross section
and mass of the lightest neutralino. Supersymmetry is capable of providing
a dark matter candidate with a present abundance consistent with WMAP and other experiments.  

Many methods have been proposed to search for evidence of supersymmetric dark matter.  These include experiments which hope to measure the recoil of dark matter particles elastically scattering off of a detector (direct searches) \cite{direct}, experiments which hope to observe the products of dark matter annihilation (indirect searches) and, of course, collider experiments \cite{collider}. Indirect searches include searches for neutrinos \cite{indirectneutrino}, gamma-rays \cite{indirectgamma}, anti-protons \cite{antiprotons} and positrons \cite{positrons,posbaltz}.

In 1994 and 1995, the High-Energy Antimatter Telescope (HEAT) observed
a flux of cosmic positrons well in excess of the predicted rate, peaking around $\sim10\,$ GeV \cite{heat1995}. This result was confirmed by another HEAT flight in 2000 \cite{heat2000,heat}. Although the source of these positrons is not known, it has been suggested in numerous publications that this signal could be the product of dark matter annihilations, particularly within the context of supersymmetry \cite{positrons,posbaltz}.

If the dark matter is evenly distributed in our local region (within a few kpc), the rate of annihilations will be insufficient to produce the observed excess. It has been suggested, however, that if sufficient clumping were present in the galactic halo, the rate at which such particles annihilate could be enhanced enough to accommodate the data. At this time, we have very little information regarding the presence of any dark matter substructure in our local region. This makes any predictions of the positron flux very difficult to make. With sophisticated numerical simulations and analytical models of galactic substructure, however, such predictions can be made on a statistical basis. In this article, we attempt to discern whether it is possible for supersymmetry to provide a flux of positrons sufficient to account for the observed excess, and if so, how much local dark matter substructure would be needed and what is the probability that the required amount of local substructure is present. We find that, although such an explanation is possible, the probability of the necessary local substructure being present is very small.

\section{Positrons From Dark Matter Annihilations}

Positrons can be produced in several neutralino annihilation modes. For example, they can result from the decay of gauge bosons produced in the interactions $\chi^0 \chi^0 \rightarrow ZZ$ or $\chi^0 \chi^0 \rightarrow W^{+}W^{-}$, producing positrons of energy $\sim m_{\chi^0}/2$. A continuum of positrons, extending to much lower energies, can also be produced in the cascades of particles produced in annihilations including fermions, Higgs bosons and gauge bosons. The spectrum of positrons produced in neutralino annihilations can vary significantly depending on the mass and annihilation modes of the LSP.

If the lightest neutralino is lighter than the $W^{\pm}$ and $Z$ bosons, their annihilations will be dominated by $\chi^0 \chi^0 \rightarrow b \bar{b}$ with a small (typically a few percent) contribution of $\chi^0 \chi^0 \rightarrow \tau^{+} \tau^{-}$. Assuming that these annihilations are dominated by bottom quarks, the spectrum of positrons produced will depend only on the mass of the LSP. If the LSP is heavier, the annihilation products can be more complicated, often dominated by several modes including $\chi \chi \rightarrow W^{+}W^{-}$, $\chi^0 \chi^0 \rightarrow ZZ$ or $\chi^0 \chi^0 \rightarrow t \bar{t}$ as well as $\chi^0 \chi^0 \rightarrow b \bar{b}$ and $\chi^0 \chi^0 \rightarrow \tau^{+} \tau^{-}$. Also, a small fraction of annihilations can directly produce an electron-positron pair, $\chi^0 \chi^0 \rightarrow e^{+} e^{-}$, although this occurs very rarely, making its impact generally negligible. In our calculations, the spectrum from each annihilation mode, including cascading, is calculated using PYTHIA \cite{pythia}, as it is implemented in the DarkSusy package \cite{darksusy}.

\begin{figure}[thb]
\vbox{\kern2.4in\includegraphics{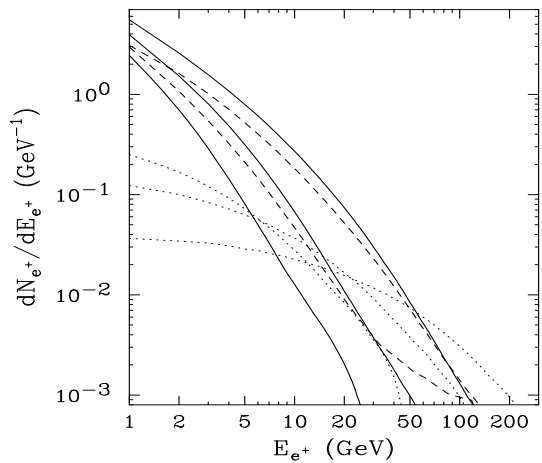}}

\caption{The positron spectrum from neutralino annihilations for the most important annihilation modes. Solid lines represent the positron spectrum, per annihilation, for $\chi^0 \chi^0 \rightarrow b \bar{b}$, for LSPs with masses of 50, 150 and 600 GeV. The dotted lines are the same, but from the process $\chi^0 \chi^0 \rightarrow \tau^{+} \tau^{-}$. Dashed lines represent positrons from the process $\chi \chi \rightarrow W^{+}W^{-}$ for LSPs with masses of 150 and 600 GeV. The spectrum from $\chi \chi \rightarrow ZZ$ is very similar. The positron spectrum from WIMP annihilations will be the weighted sum of the spectra over the annihilation modes, such as those shown here. See text for more details.}
\label{spectrum}
\end{figure}

In figure~\ref{spectrum}, we show the positron spectrum from neutralino annihilations for the most important annihilation modes. Solid lines represent the positron spectrum, per annihilation, for $\chi^0 \chi^0 \rightarrow b \bar{b}$, for LSPs with masses of 50, 150 and 600 GeV. The dotted lines are the same, but from the process $\chi^0 \chi^0 \rightarrow \tau^{+} \tau^{-}$. Gaugino-like annihilations typically produce a spectrum which is dominated by $b \bar{b}$ at low energies, with contributions from $\tau^{+} \tau^{-}$ only becoming important at energies of about half the LSP mass and above.

For neutralinos with a substantial higgsino component, annihilations to gauge boson will often dominate. Dashed lines represent positrons from the process $\chi \chi \rightarrow W^{+}W^{-}$ for LSPs with masses of 150 and 600 GeV. The spectrum from $\chi \chi \rightarrow ZZ$ is very similar. Notice that the positron spectrum via gauge bosons is slightly harder than from $b \bar{b}$.

\section{Positron Propagation}

We use a standard diffusion model to calculate the effect of propagation on the observed positron spectrum. In this model, positrons, being charged particles, move under the influence of interstellar magnetic fields. The galactic magnetic field is tangled, resulting in motion which is well described as a random walk. During this motion, positrons lose energy via inverse Compton and synchrotron processes. 

The diffusion-loss equation describing this process is given by

\begin{eqnarray}
\frac{\partial}{\partial t}\frac{dn_{e^{+}}}{dE_{e^{+}}} &=& \vec{\bigtriangledown} \cdot \bigg[K(E_{e^{+}},\vec{x})  \vec{\bigtriangledown} \frac{dn_{e^{+}}}{dE_{e^{+}}} \bigg] \nonumber\\
&+& \frac{\partial}{\partial E_{e^{+}}} \bigg[b(E_{e^{+}},\vec{x})\frac{dn_{e^{+}}}{dE_{e^{+}}}  \bigg] + Q(E_{e^{+}},\vec{x}),
\label{dif}
\end{eqnarray}
where $dn_{e^{+}}/dE_{e^{+}}$ is the number density of positrons per unit energy, $K(E_{e^{+}},\vec{x})$ is the diffusion constant, $b(E_{e^{+}},\vec{x})$ is the rate of energy loss and $Q(E_{e^{+}},\vec{x})$ is the source term.

We parameterize the diffusion constant \cite{diffusion} and rate of energy loss by

\begin{equation}
K(E_{e^{+}}) = 3 \times 10^{27} \bigg[3^{0.6} + E_{e^{+}}^{0.6} \bigg] \,\rm{cm}^2 \, \rm{s}^{-1}
\label{k}
\end{equation}
and
\begin{equation}
b(E_{e^{+}}) = 10^{-16} E_{e^{+}}^2 \,\, \rm{s}^{-1},
\label{b}
\end{equation}
respectively. $b(E_{e^{+}})$ is the result of inverse Compton scattering on both starlight and the cosmic microwave background \cite{lossrate}. The diffusion parameters are constrained from analyzing stable nuclei in cosmic rays (primarily by fitting the boron to carbon ratio) \cite{L}.

In equations~\ref{k} and~\ref{b}, we have dropped the dependence on location.  We treat these as constant within the diffusion zone. For the diffusion zone, we consider a slab of thickness $2L$, where $L$ is chosen to be 4 kpc, the best fit to observations \cite{diffusion,L}. The radius of the slab is unimportant, as it is larger than the distances which positrons can propagate at these energies. Outside of the diffusion zone, we drop the positron density to zero (free escape boundary conditions).

We solve equation~\ref{dif} following the procedure described in Ref.~\cite{posbaltz}. For detailed descriptions of two zone diffusion models, see Ref.~\cite{posbaltz,L,2zonediffusion}.

\section{Substructure}
\label{subsection}

Given the shape of the primordial power spectrum for CDM fluctuations,
dark matter halos are expected to form hierarchically from the merging
of smaller bound systems, the cores of which survive as dense substructure 
in present-day halos. This merger process has been studied extensively
using numerical $N$-body simulations. While simulations should provide 
an accurate picture of halo formation through merging, they cannot 
easily track the evolution of substructure in the densest regions
of halos. Even the highest resolution simulations show little or no substructure
in their innermost regions, but this is probably due to numerical effects rather 
than physical ones. To avoid underestimating the number of dark matter clumps 
that might survive in the solar neighbourhood, we will use a recently developed 
semi-analytic model of halo formation \cite{tbprep}. The model treats 
the evolution of dark matter lumps in a way that is less sensitive to their mass or the mean density of their environment. Comparisons with simulations suggest 
that if anything, this model may overestimate the amount of substructure in halos \cite{tbprep}.

We will outline the basic features of this model here; for detailed information, see Ref.~\cite{tbprep}. The
model generates sets of random but representative merger histories for
dark matter halos, using the merger-tree algorithm of Ref.~\cite{somerville}. The merger trees are `pruned' \cite{tbprep} to determine whether halos merging with the main system
contribute a single lump or several lumps (corresponding to their own undigested 
substructure) to the substructure of the main system. 
The dynamical evolution of individual lumps 
is determined using the analytic model of satellite dynamics developed in Ref.~\cite{tb2001}, which includes the effects of dynamical friction, 
tidal mass loss and tidal shock-heating. The background mass distribution 
is taken to be the sum of a spherical dark matter halo (with a cosmological density
profile), a spherical bulge component and an exponential disk. 
The dark matter halo grows in mass according to 
its merger history, and changes in concentration following the 
relations proposed in Ref.~\cite{eke}. The growth of the baryonic
components is modelled as described in Ref.~\cite{islam} 
in order to produce present-day systems with roughly the properties of
the Milky Way. 

We run this model using a large set of input merger trees
with a mass resolution of $5\times 10^7 M_{\odot}$, as well as a smaller 
number of trees with a resolution of $1\times 10^7 M_{\odot}$, in order 
to test for convergence in our results. Within each model halo, we then 
select every surviving lump in the solar neighbourhood 
(7 kpc $ < R < $ 10 kpc). Given the average amount of 
stripping experienced by these systems, our results for this region
are complete down to roughly $3 \times 10^6 M_{\odot}$ and 
$6 \times 10^5 M_{\odot}$ for the low-resolution and high-resolution 
models, respectively. For each of these systems, 
we calculate a quantity proportional to the annihilation rate, $f \times M^2_{\rm{clump}}/V_{\rm{clump}}$,
assuming they start out with an NFW density profile with a concentration
given by the ENS concentration relations \cite{eke}, and that stripping changes their density profile
as described in Ref.~\cite{hayashi}, and the value of $f$
as described in \cite{ts2003}. Here, $f$ is defined as $\int (\rho^2/\bar{\rho}^2) dV$.

We note that assuming a density profile with a central slope close 
to $r^{-1.5}$ for the subhalos would result in annihilation rates up 
to 20 times larger \cite{ts2003}, but a profile this steep 
is disfavoured by the most recent numerical results \cite{nomoore}.

\begin{figure}[!ht]
\vbox{\kern2.4in\includegraphics{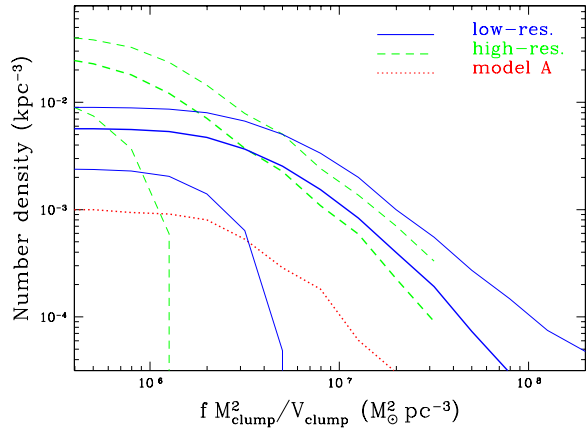}}
\caption{The number density of dark matter clumps between 7 and 10
kpc from the galactic center, as a function of the minimum value of 
$f \times M^2_{\rm{clump}}/V_{\rm{clump}}$ considered. The thick solid line shows
the results for merger trees complete down to $\sim 3\times 10^6 M_{\odot}$,
while the thick dashed line shows the results for higher-resolution trees
complete down to $\sim 6\times 10^5 M_{\odot}$. The thick dotted line shows 
results for a different disruption efficiency (see text). The thin lines
show the $\pm 1$-$\sigma$ halo-to-halo variation.}
\label{fig:sub}
\end{figure}

Figure \ref{fig:sub} shows the number density of dark matter clumps between 
7 and 10 kpc from the galactic center, as a function of the minimum value of 
$f \times M^2_{\rm{clump}}/V_{\rm{clump}}$ considered. The solid lines show the
results for the low-resolution trees (the thick line indicates the average,
and the thin lines show the 1-$\sigma$ halo-to-halo variation), while the
dashed lines show the results for the high-resolution trees. The two
agree to within a fraction of the halo-to-halo scatter, for values of $f \times M^2_{\rm{clump}}/V_{\rm{clump}}$ above their respective resolutions, indicating 
our estimate is reasonably independent of merger-tree resolution. 
Clearly the halo-to-halo scatter is very large; some systems have 
almost no clumps close to the solar radius, for instance. 
There is a fairly well-defined upper envelope to the distribution, 
however. For example, for values of $f \times M^2_{\rm{clump}}/V_{\rm{clump}} > 2\times 10^7 M^2_{\odot}\,\rm{pc}^{-3}$, no more than about $10^{-3}$ clumps per cubic kpc are predicted. Furthermore, we have reason to believe this is a very conservative
upper limit. Not only do simulations find less substructure locally,
as mentioned previously, but even within the semi-analytic model there
is some uncertainty as to whether this many systems really survive.
We have assumed that subhalos on general orbits can survive stripping 
down to one tenth of the critical radius defined by Ref.~\cite{hayashi}
for circular orbits, by which point they have typically lost more than 99\% 
of their original mass (`model B' in Ref.~\cite{tbprep}). If we make 
the less extreme assumption that systems are disrupted once they have
been stripped down to half their critical radius (97--98\% mass loss),
then far fewer systems survive in the solar neighbourhood, as indicated
by the dotted line in figure \ref{fig:sub}.

\section{Positron Spectra From Dark Substructure}

We will now consider a single clump of dark matter particles. We justify this assumption by pointing out that the probability of having a single clump within a few kpc which contributes some particular amount to the annihilation rate is considerably larger than the probability of having multiple clumps which sum to the same contribution. This can be observed in figure~\ref{fig:sub}. 

The annihilation rate in a dark matter clump is dependent on the clump's mass and density profile as well as the dark matter particle's mass and annihilation cross section. Treating the effect of these quantities as a free parameter used to normalize the flux of positrons to observations, we can assess if the observed spectrum can be fit by such a scenario.

We will consider five representative supersymmetry scenarios. First, a light (50 GeV) neutralino which annihilates 96\% of the time into $b \bar{b}$ and 4\% to $\tau^+ \tau^-$. Such a particle could be gaugino-like or higgsino-like, as below the gauge boson masses, these modes dominate for either case. 

Second, we consider two cases for an intermediate mass (150 GeV) neutralino. One which annihilates as described in the previous case, and another which annihilates to gauge bosons (58\% to $W^+ W^-$ and 42\% to $ZZ$). These neutralinos are typically gaugino-like and higgsino-like, respectively.

Finally, we consider two cases of heavy (600 GeV) neutralinos. The first (gaugino-like) annihilates mostly (87\%) to $b \bar{b}$ and to $\tau^+ \tau^-$ or $t \bar{t}$ the remaining times. The second (higgsino-like) annihilates 65\% of the time to gauge bosons, 19\% to $\tau^+ \tau^-$ and 14\% to higgs bosons.

Although these five cases do not completely encompass the very wide array of characteristics neutralinos may have, they do describe effective benchmarks. Furthermore, the relevant characteristics of a LSP with a mixture of these annihilation modes or another mass could be readily inferred by inspecting these representative models.

The effect of propagation on the positron spectrum depends strongly on the distance from the source (the dark matter clump) and the observer. Since the tidal radius of typical clumps is typically much smaller than the distances to clumps we will consider, we can treat such a clump spatially as a point source.

 To compare to the data recorded by HEAT, rather than simply the positron flux, we consider the ratio of the positron flux to the combined positron and electron fluxes, called the positron fraction. We do this by using the spectra for secondary positron, secondary electrons and primary electrons found in Ref.~\cite{secbg}. 

\begin{figure}[thb]
\vbox{\kern2.4in\includegraphics{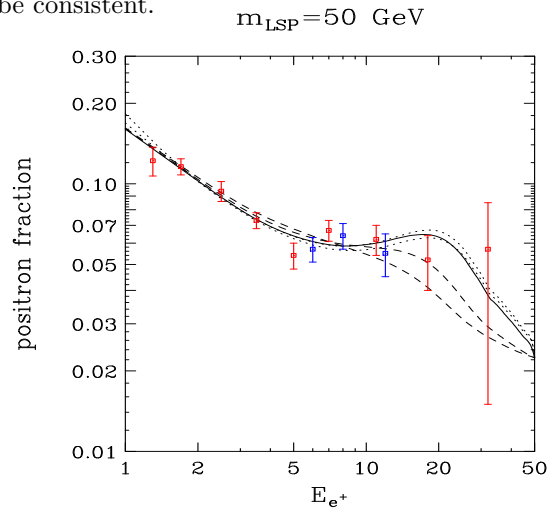}}
\caption{The positron fraction, as a function of positron energy (in GeV), for a 50 GeV neutralino which annihilates 96\% to $b \bar{b}$ and 4\% to $\tau^+ \tau^-$. The solid line represents the distance to the dark matter clump at which the predicted spectrum best fits the data. Dotted and dashed lines represent the spectra for a source at (1 and 2$\sigma$) distances less than and greater than found for the best fit, respectively. The normalization was considered to be a free parameter. The error bars shown are for the 94-95 and 2000 HEAT flights.}
\label{fit1}
\end{figure}

\begin{figure}[thb]
\vbox{\kern2.4in\includegraphics{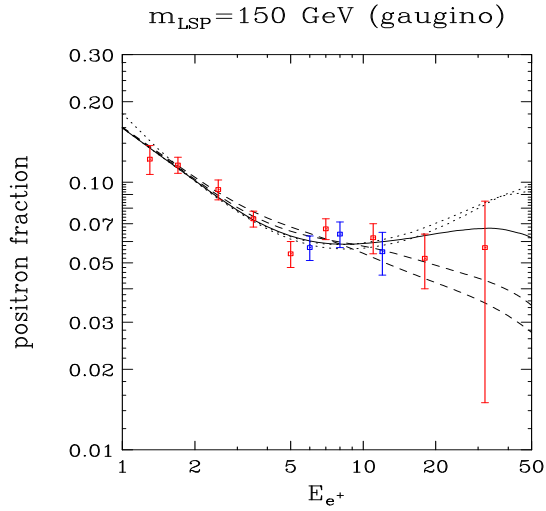}}

\caption{The predicted positron fraction, as a function of positron energy (in GeV), for a 150 GeV neutralino which annihilates 96\% to $b \bar{b}$ and 4\% to $\tau^+ \tau^-$. Otherwise, the same as in figure~\ref{fit1}.}
\label{fit2}
\end{figure}

\begin{figure}[thb]
\vbox{\kern2.4in\includegraphics{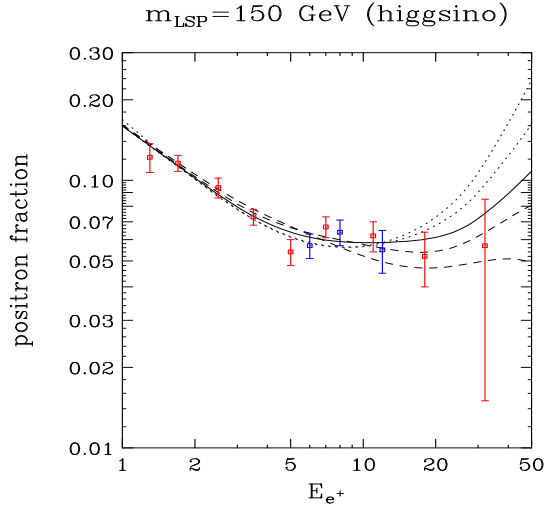}}

\caption{The predicted positron fraction, as a function of positron energy (in GeV), for a 150 GeV neutralino which annihilates 58\% to $W^+ W^-$ and 42\% to $ZZ$. Otherwise, the same as in figure~\ref{fit1}.}
\label{fit3}
\end{figure}

\begin{figure}[thb]
\vbox{\kern2.4in\includegraphics{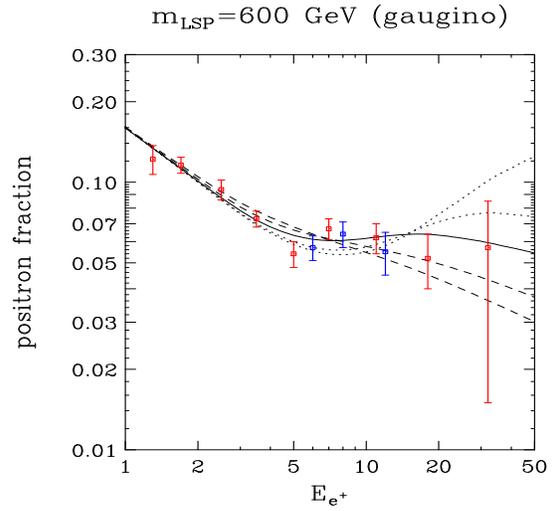}}

\caption{The predicted positron fraction, as a function of positron energy (in GeV), for a 600 GeV neutralino which annihilates 87\% to $b \bar{b}$ and to $\tau^+ \tau^-$ or $t \bar{t}$ the remaining times. Otherwise, the same as in figure~\ref{fit1}.}
\label{fit4}
\end{figure}

\begin{figure}[thb]
\vbox{\kern2.4in\includegraphics{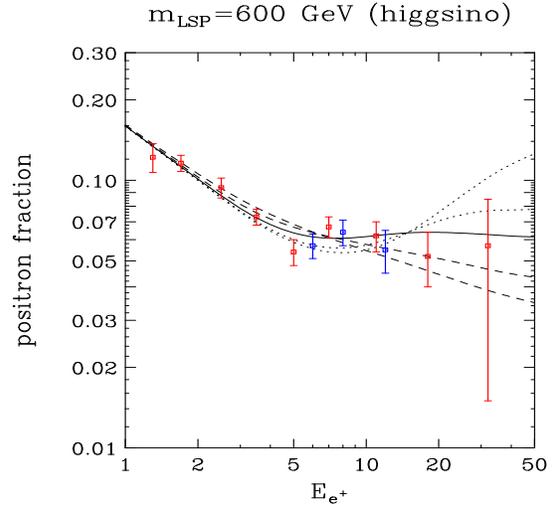}}

\caption{The predicted positron fraction, as a function of positron energy (in GeV), for a 600 GeV neutralino which annihilates 65\% to gauge bosons, 19\% $\tau^+ \tau^-$ and 14\% to Higgs bosons. Otherwise, the same as in figure~\ref{fit1}.}
\label{fit5}
\end{figure}

Figures~\ref{fit1}-\ref{fit5} show the positron fraction, as a function of positron energy, for each of the five representative supersymmetry scenarios described above. We varied the distance to the source (the dark matter clump) to determine the effect on the positron spectrum. In each figure, the solid line represents the distance at which the predicted spectrum is best fit to the data. In all five cases, we found very good fits to the HEAT observations.  

Dotted and dashed lines represent the spectra for a source at (1 and 2$\sigma$) distances less than and greater than found for the best fit, respectively. For these lines, the $\chi^2$ is larger by 1 and 4, respectively (1$\sigma$ and 2$\sigma$). The normalization was considered to be a free parameter. The predicted spectrum is compared to the error bars of the 94-95 and 2000 HEAT data. These results are summarized in tables I and II. 

Note that our results are quite different from that of other collaborations which treat the spatial distribution of WIMP annihilations as constant. Introducing a spatial distribution of clumps allows the data to be well fit by our predictions for SUSY models which would not otherwise be consistent.

\vspace{0.2cm}

\vspace{-0.3cm}
 \begin{table}[!ht]
 \label{table1}
 \hspace{1.5cm}
 \begin{tabular} {c c c c c} 
 SUSY Model & Best-Fit & 1$\sigma$ & 2$\sigma$ & \\
 \hline \hline
 $m_{\chi^0}$=50 GeV (Case 1) & 0.27 kpc & 0.21-0.68 kpc & 0.20-1.1 kpc & \\
 \hline
 $m_{\chi^0}$=150 GeV (Case 2) & 0.42 kpc & 0.23-0.85 kpc & 0.19-1.3 kpc & \\
 \hline
 $m_{\chi^0}$=150 GeV (Case 3) & 0.62 kpc & 0.20-1.1 kpc & 0.20-1.6 kpc & \\
 \hline
 $m_{\chi^0}$=600 GeV (Case 4) & 0.82 kpc & 0.44-1.4 kpc & 0.20-2.1 kpc & \\
 \hline
 $m_{\chi^0}$=600 GeV (Case 5) & 0.87 kpc & 0.49-1.5 kpc & 0.23-2.2 kpc & \\
 \end{tabular}
 \caption{The distances to a dark matter clump found to best fit the observed positron spectrum for the five supersymmetry models described in the text. One and two sigma error bars are also given. Cases 1, 2 and 4 correspond to models dominated by annihilation to $b \bar{b}$ (with a small fraction annihilating to $\tau^- \tau^+$ or, in the heaviest case, $t \bar{t}$). Cases 3 and 5 correspond to models which annihilate dominantly into gauge bosons (with, in the heavy case, a small fraction annihilating to $b \bar{b}$, $\tau^- \tau^+$ or $t \bar{t}$).}
 \end{table}


\vspace{-0.3cm}
 \begin{table}[!ht]
 \label{table2}
 \hspace{1.5cm}
 \begin{tabular} {c c c c c} 
 SUSY Model &  Using Best-Fit Dist. &  Using 2$\sigma$ Dist.& \\
 \hline \hline
 $m_{\chi^0}$=50 GeV (Case 1) & $2.3 \times 10^{37}$ &  $2.0-5.0 \times 10^{37}$  & \\
 \hline
 $m_{\chi^0}$=150 GeV (Case 2) & $5.0 \times 10^{36}$ &  $3.9-13. \times 10^{36}$  & \\   
 \hline
 $m_{\chi^0}$=150 GeV (Case 3) & $9.2 \times 10^{36}$ &  $6.2-26. \times 10^{36}$  & \\
 \hline
 $m_{\chi^0}$=600 GeV (Case 4) & $2.1 \times 10^{36}$ &  $0.9-8.9 \times 10^{36}$  & \\
 \hline
 $m_{\chi^0}$=600 GeV (Case 5) & $2.8 \times 10^{36}$ &  $1.0-12. \times 10^{36}$  & \\
 \end{tabular}
 \caption{The rate of neutralino annihilations (per second) in a dark matter clump required to account for the observed positron excess, using the distances in the previous table. Again, the five supersymmetry models which have been described in the text are considered. Results are shown for a clump the best-fit distance from Earth, and for a clump at a distance corresponding to 2$\sigma$ from the best fit.}
 \end{table}
\section{Assessment}

The results shown in figures~\ref{fit1}-\ref{fit5} and tables I and II can be used to help assess the size and density profile of a dark matter clump required to account for the observed positron excess. The annihilation rate in a clump is a function of the annihilation cross section, the LSP mass and clump characteristics (mass and density profile):

\begin{equation}
R = \frac{1}{2}\frac{<\sigma v>}{m_{\rm{LSP}}^2} \frac{M_{\rm{clump}}^2}{V_{\rm{clump}}} \times  f,
\end{equation}
where $<\sigma v>$ is the low-velocity neutralino annihilation cross section times velocity, $m_{\rm{LSP}}$ is the neutralino mass, $M_{\rm{clump}}$ is the mass of the clump, $V_{\rm{clump}}$ is the volume of the clump, and $f$ is the enhancement from uneven distributions of mass within the clump, defined in section \ref{subsection}.

The first factor in the expression for the annihilation rate depends on particle physics. To ascertain how large this factor ($<\sigma v>/m_{\rm{LSP}}^2$) could be, we ran a Monte Carlo, randomly selecting parameters of the MSSM. This method found several thousand SUSY models which respected all collider constraints and provided a neutralino LSP with a relic density in the range measured by WMAP ($0.095 < \Omega_{\rm{\chi^0}} h^2 < 0.129$). The models were selected by randomly varying values of $M_2$ between 10 and 10,000 GeV, $\vert \mu \vert$ between 10 and 10,000 GeV (with either sign), the mass of the pseudoscalar Higgs, $m_A$, between 30 and 1000 GeV, $\tan \beta$ between 1 and 60, and the squark and slepton masses between 30 and 10,000 GeV. The trilinear couplings, $A_t$ and $A_b$, were varied between 30 and 10,000 GeV, as well. $M_1$ and $M_3$ were determined by the GUT relations. Of the models our Monte Carlo produced, none with LSP masses in the range of 30-70 GeV had values of $<\sigma v> / m_{\rm{LSP}}$ larger than about $2 \times 10^{-29} \,\,\rm{cm}^3 \, \rm{s}^{-1} \, \rm{GeV}^{-2}$. For the other four characteristic SUSY scenarios we are considering throughout this paper, we find upper values for this factor of $4 \times 10^{-30}$ (150 GeV, gaugino), $2 \times 10^{-30}$ (150 GeV, higgsino), $1 \times 10^{-31}$ (600 GeV, gaugino) and $1 \times 10^{-31}$ (600 GeV, higgsino) $\,\,\rm{cm}^3 \, \rm{s}^{-1} \, \rm{GeV}^{-2}$. We will use these values for $<\sigma v>/m_{\rm{LSP}}^2$ for the remainer of our calculation in an effort to be optimistic.


For a light LSP ($m_{\chi^0} < m_{W^{\pm}}$), annihilations will be dominated by $b \bar{b}$ except for a small fraction of models which have very small annihilation cross section and are, therefore, not of interest to us. As we said before, for a 50 GeV neutralino, $<\sigma v>/m_{\rm{LSP}}^2$ can be as large as $\sim 2 \times 10^{-29}\,\,\rm{cm}^{3}\,\rm{s}^{-1}\, \rm{GeV}^{-2}$. Combining this with the result of table II, this leads to the requirement of a clump with $f \times M_{\rm{clump}}^2 / V_{\rm{clump}} \simeq 2.3 \times 10^7 \, M_{\odot}^2/\rm{pc}^3$ using the best fit distance and $2.0-5.0 \times 10^7 \, M_{\odot}^2/\rm{pc}^3$ using the distances corresponding to a $2\sigma$ fit. These results and the results for our other supersymmetry benchmarks are shown in table III. 


Now that we have determined the nature of the clump required to account for the observed positron excess, we can begin to estimate the probability of such a clump being present. For example, considering a 50 GeV neutralino, if we require a clump within the best-fit distance of 0.27 kpc and with at least $f \times M_{\rm{clump}}^2 / V_{\rm{clump}} \simeq 2.3 \times 10^7 \,\, M_{\odot}^2/\rm{pc}^3$, using the results of figure~\ref{fig:sub}, we see that the probability of this occuring is approximately the number density of such clumps multiplied by the volume within this distance, $(2-3 \times 10^{-4})\, (4 \pi \times 0.27^3 /3) \simeq 2.5 \times 10^{-5}$, or one in 40,000. These results and the results for other SUSY models are shown in table IV.

Note that if non-thermal processes are responsible for the observed density of dark matter \cite{nonthermal}, the neutralino annihilation cross section may be considerably larger and our estimates inaccurate.




\vspace{0.0cm}
 \begin{table}
 \label{table3}
 \hspace{1.5cm}
 \begin{tabular} {c c c c} 
 SUSY Model &  Using Best-Fit Dist. &  Using 2$\sigma$ Dist.& \\
 \hline \hline
 $m_{\chi^0}$=50 GeV (Case 1) & $2.3 \times 10^7$  &  $2.0-5.0 \times 10^7$     & \\
 \hline
 $m_{\chi^0}$=150 GeV (Case 2) & $2.1 \times 10^7$   &  $1.7-5.5 \times 10^7$         & \\
 \hline
 $m_{\chi^0}$=150 GeV (Case 3) &  $9.2 \times 10^7$   &  $6.2-26. \times 10^7$      & \\
 \hline
 $m_{\chi^0}$=600 GeV (Case 4) &   $4.2 \times 10^8$   &  $1.8-18. \times 10^8$     & \\
 \hline
 $m_{\chi^0}$=600 GeV (Case 5) & $5.6 \times 10^8$  &  $2.0-24. \times 10^8$    & \\
 \end{tabular}
 \caption{The values of $f \times M_{\rm{clump}}^2 / V_{\rm{clump}}$, in units of $M_{\odot}^2/\rm{pc}^3$, required from a dark matter clump to account for the observed positron excess.  The five supersymmetry models described in the text are again used. These values are given for the distance to the clump which results in the best fit spectrum as well as for the distances corresponding to the two sigma error bars around the best fit. The values of $<\sigma v> / m_{\rm{LSP}}$ used are fixed to optimistic values of $2 \times 10^{-29}$, $4 \times 10^{-30}$, $2 \times 10^{-30}$, $1 \times 10^{-31}$ and $1 \times 10^{-31} \,\,\rm{cm}^3 \, \rm{s}^{-1} \, \rm{GeV}^{-2}$, for each of the five supersymmetry models used, respectively. Cases 1, 2 and 4 correspond to models dominated by annihilation to $b \bar{b}$ (with a small fraction annihilating to $\tau^- \tau^+$ or, in the heaviest case, $t \bar{t}$). Cases 3 and 5 correspond to models which annihilate dominately into gauge bosons (with, in the heavy case, a small fraction annihilating to $b \bar{b}$, $\tau^- \tau^+$ or $t \bar{t}$).}
 \end{table}
\vspace{0.0cm}
 \begin{table}
 \label{table4}
 \hspace{1.5cm}
 \begin{tabular} {c c c c} 
 SUSY Model &  Using Best-Fit Dist. &  Using 2$\sigma$ Dist.& \\
 \hline \hline
 $m_{\chi^0}$=50 GeV (Case 1) & $3 \times 10^{-5}$  & $5 \times 10^{-4}$ & \\
 \hline
 $m_{\chi^0}$=150 GeV (Case 2) & $1 \times 10^{-4}$   &  $8 \times 10^{-4}$ & \\
 \hline
 $m_{\chi^0}$=150 GeV (Case 3) &  $2 \times 10^{-5}$   &  $ 2 \times 10^{-4}$     & \\
 \hline
 $m_{\chi^0}$=600 GeV (Case 4) &   $\lsim 10^{-5}$   &  $\sim 10^{-4}-10^{-5}$   & \\
 \hline
 $m_{\chi^0}$=600 GeV (Case 5) & $\lsim 10^{-5}$  &   $\sim  10^{-4}-10^{-5}$      & \\
 \end{tabular}
 \caption{The probability of a clump with sufficient $f \times M_{\rm{clump}}^2 / V_{\rm{clump}}$ within the best fit and $2\sigma$ distances. Other details are the same as in table III. The most optimistic values for cross sections and substructure number densities were used.}
 \end{table}

\section{Conclusions}

In each characteristic supersymmetry model considered, we found that (up to normalization) the spectrum of positrons observed by HEAT could be well fit by a dark matter clump at an unknown distance acting as a point source of positrons. Furthermore, by treating dark matter clumps as point sources rather than using an even distribution, we can estimate the distance to the source clump for a given neutralino mass and dominating annihilation modes. Such distances are typically 0.2 to 2.0 kiloparsecs, considerably larger than the average size of the core or tidal radius of such clumps. Thus the point-like approximation for annihilations is justified. In the unlikely case that the solar system resides inside of such a clump, the spectrum will be quite sharply rising at high energies, fitting the data poorly, as can be inferred from figures 3-7.

Once a source distance is estimated, we can determine the annihilation rate which must occur in the clump to properly normalize the positron flux. This rate is the result of both the characteristics of the neutralino ($<\sigma v> / m_{\rm{LSP}}^2$) and the clump ($f \times M_{\rm{clump}}^2 / V_{\rm{clump}}$). By considering a model of galactic substructure which can predict the number density of dark matter clumps, we can use the required annihilation rate to estimate the probability of a clump being present which is sufficient to produce the observed positron excess for a given supersymmetry model. It is likely that our substructure model overestimates the number of clumps. With this in mind, we consider this as a rough upper limit.

We find that, using optimistic neutralino annihilation cross sections and annihilation modes, the probability of local dark matter substructure being sufficient to provide the observed positron excess is quite small, on the order of $10^{-4}$ or less. Given this, we conclude that although it would not be impossible for annihilating supersymmetric dark matter to produce the observed positron excess, it is a very unlikely and unnatural solution.

\vspace{0.5cm}

{\it Acknowledgments}: DH and JT are supported by the Leverhulme Trust.


\vskip -0.5cm
 
\begin{thebibliography}{99} 



\bibitem{structure}
K.~Abazajian {\it et al.}  [SDSS Collaboration],
arXiv:astro-ph/0305492;
K.G.~Begeman, A.H.~Broeils, R.H.~Sanders,
Mon. Not. R. Astr. Soc. {\bf 249}, 523 (1991);
M.~Davis, G.~Efstathiou, C.~S.~Frenk and S.~D.~White,
Astrophys.\ J.\  {\bf 292}, 371 (1985).

%
\bibitem{supernovae}
S.~Perlmutter {\it et al.}  [Supernova Cosmology Project Collaboration],
Astrophys.\ J.\  {\bf 517}, 565 (1999)
[astro-ph/9812133].
%
\bibitem{cmb}
P.~de Bernardis {\it et al.}, Nature {\bf 404}, 955 (2000)
[astro-ph/0004404];
S.~Hanany {\it et al.},
Astrophys.\ J.\  {\bf 545}, L5 (2000)
[arXiv:astro-ph/0005123];
A.~Balbi {\it et al.},
Astrophys.\ J.\  {\bf 545}, L1 (2000)
[Erratum-ibid.\  {\bf 558}, L145 (2001)];
C.~B.~Netterfield {\it et al.}  [Boomerang Collaboration],
Astrophys.\ J.\  {\bf 571}, 604 (2002).

\bibitem{wmap}
C.~L.~Bennett {\it et al.},
arXiv:astro-ph/0302207.

\bibitem{nonbaryonic}
S.~Burles, K.~M.~Nollett, J.~N.~Truran and M.~S.~Turner,
Phys.\ Rev.\ Lett.\  {\bf 82}, 4176 (1999);
M.~Fukugita, C.~J.~Hogan and P.~J.~Peebles, 
Astrophys.~J. {\bf 503}, 518 (1998);
S.~Burles, K.M.~Nollett, J.N.~Truran and M.S.~Turner,
Phys.\ Rev.\ Lett.\ {\bf 82}, 4176 (1999)
astro-ph/9901157.



\bibitem{susylsp}
H.~Goldberg, Phys.\ Rev.\ Lett.\ {\bf 50}, 1419 (1983);
%
J.~Ellis, J.~S.~Hagelin, D.~V.~Nanopoulos, K.~Olive and M.~Srednicki,
Nucl.\ Phys.\ B {\bf 238}, 453 (1984).


\bibitem{rparity}
 S.~Weinberg,
  Phys.\ Rev.\ D {\bf 26}, 287 (1982);
 L.~J.~Hall and M.~Suzuki,
  Nucl.\ Phys.\ B {\bf 231}, 419 (1984);
 B.~C.~Allanach, A.~Dedes and H.~K.~Dreiner,
  Phys.\ Rev.\ D {\bf 60}, 075014 (1999).







\bibitem{direct}
A.~Drukier and L.~Stodolsky,
Phys.\ Rev.\ D {\bf 30}, 2295 (1984);
M.~W.~Goodman and E.~Witten,
Phys.\ Rev.\ D {\bf 31}, 3059 (1985);
A.~Bottino, V.~de Alfaro, N.~Fornengo, S.~Mignola and S.~Scopel,
Astropart.\ Phys.\  {\bf 2}, 77 (1994);
H.~Baer and M.~Brhlik,
Phys.\ Rev.\ D {\bf 57}, 567 (1998).



\bibitem{collider}
 R.~Barate {\it et al.}  [ALEPH Collaboration],
  Phys.\ Lett.\ B {\bf 499}, 67 (2001);
 G.~Abbiendi {\it et al.}  [OPAL Collaboration],
  Eur.\ Phys.\ J.\ C {\bf 14}, 51 (2000);
 M.~Acciarri {\it et al.}  [L3 Collaboration],
  Phys.\ Lett.\ B {\bf 471}, 280 (1999);
 A.~Heister {\it et al.}  [ALEPH Collaboration],
  Phys.\ Lett.\ B {\bf 526}, 206 (2002);
 D.~Acosta {\it et al.}  [CDF Collaboration],
  Phys.\ Rev.\ D {\bf 65}, 091102 (2002);
 F.~Abe {\it et al.}  [CDF Collaboration],
  Phys.\ Rev.\ D {\bf 56}, 1357 (1997);
 B.~Abbott {\it et al.}  [D0 Collaboration],
  Phys.\ Rev.\ D {\bf 60}, 031101 (1999);
 B.~Abbott {\it et al.}  [D0 Collaboration],
  Phys.\ Rev.\ D {\bf 63}, 091102 (2001);
 S.~Abel {\it et al.}  [SUGRA Working Group Collaboration],
  hep-ph/0003154.



\bibitem{indirectneutrino}
J.~Silk, K.~Olive and M.~Srednicki,
Phys.\ Rev.\ Lett.\ {\bf 55}, 257 (1985);
J.~S.~Hagelin, K.~W.~Ng, K.~A.~Olive, 
Phys.\ Lett.\ B {\bf 180}, 375 (1986); 
F.~Halzen, T.~Stelzer and M.~Kamionkowski,
Phys.\ Rev.\ D {\bf 45}, 4439 (1992);
J.~L.~Feng, K.~T.~Matchev and F.~Wilczek,
Phys.\ Rev.\ D {\bf 63}, 045024 (2001);
V.~D.~Barger, F.~Halzen, D.~Hooper and C.~Kao,
Phys.\ Rev.\ D {\bf 65}, 075022 (2002);
J.~L.~Feng, K.~T.~Matchev and F.~Wilczek,
Phys.\ Rev.\ D {\bf 63}, 045024 (2001);
V.~Berezinsky, A.~Bottino, J.~Ellis, N.~Fornengo, G.~Mignola and S.~Scopel,
Astropart.\ Phys.\ {\bf 5}, 333 (1996)
[hep-ph/9603342];
L.~Bergstrom, J.~Edsjo and P.~Gondolo,
Phys.\ Rev.\ D {\bf 55}, 1765 (1997);
Phys.\ Rev.\ D {\bf 58}, 103519 (1998);
L.~Bergstrom, J.~Edsjo and M.~Kamionkowski,
Astropart.\ Phys.\ {\bf 7}, 147 (1997);
A.~Corsetti and P.~Nath,
Int.\ J.\ Mod.\ Phys.\ A {\bf 15}, 905 (2000);
A.~E.~Faraggi, K.~A.~Olive and M.~Pospelov,
Astropart.\ Phys.\ {\bf 13}, 31 (2000);
K.~Freese,
Phys.\ Lett.\ B {\bf 167}, 295 (1986);
L.~M.~Krauss, M.~Srednicki and F.~Wilczek,
Phys.\ Rev.\ D {\bf 33}, 2079 (1986);
T.~K.~Gaisser, G.~Steigman and S.~Tilav,
Phys.\ Rev.\ D {\bf 34}, 2206 (1986).
%



\bibitem{indirectgamma}
F~ W.~Stecker and A.~J.~Tylka, 
Astrophys.\ J. 343, 169 (1989);
F.~W.~Stecker, Phys.\ Lett.\ B 201, 529 (1988);
L.~Bergstrom, J.~Edsjo and P.~Ullio,
Phys.\ Rev.\ Lett.\  {\bf 87}, 251301 (2001);
L.~Bergstrom, J.~Edsjo and C.~Gunnarsson,
Phys.\ Rev.\ D {\bf 63}, 083515 (2001);
H.~U.~Bengtsson, P.~Salati and J.~Silk,
Nucl.\ Phys.\ B {\bf 346}, 129 (1990);
V.~Berezinsky, A.~Bottino and G.~Mignola,
Phys.\ Lett.\ B {\bf 325}, 136 (1994)
[arXiv:hep-ph/9402215];
L.~Bergstrom and P.~Ullio,
Nucl.\ Phys.\ B {\bf 504}, 27 (1997)
[arXiv:hep-ph/9706232];
D.~Hooper and B.~L.~Dingus,
arXiv:astro-ph/0210617;
L.~Bergstrom, P.~Ullio and J.~H.~Buckley,
Astropart.\ Phys.\  {\bf 9}, 137 (1998).


\bibitem{antiprotons}
F.~W.~Stecker, S.~Rudaz and T.~F.~Walsh,  
Phys.\ Rev.\ Letters {\bf 55}, 2622 (1985);
S.~Rudaz and F.~W.~Stecker, 
Astrophys.\ J. 325, 16 (1988);
L.~Bergstrom, J.~Edsjo and P.~Ullio,
arXiv:astro-ph/9906034;
A.~Bottino, F.~Donato, N.~Fornengo and P.~Salati,
Phys.\ Rev.\ D {\bf 58}, 123503 (1998);
F.~Donato, N.~Fornengo, D.~Maurin, P.~Salati and R.~Taillet,
arXiv:astro-ph/0306207.

\bibitem{positrons}
G.~L.~Kane, L.~T.~Wang and J.~D.~Wells,
Phys.\ Rev.\ D {\bf 65}, 057701 (2002);
M.~Kamionkowski and M.~S.~Turner,
Phys.\ Rev.\ D {\bf 43}, 1774 (1991);
E.~A.~Baltz, J.~Edsjo, K.~Freese and P.~Gondolo,
arXiv:astro-ph/0211239;
M.~S.~Turner and F.~Wilczek,
Phys.\ Rev.\ D, {\bf 42}, 1001 (1990);
A.~J.~Tylka,
Phys.\ Rev.\ Lett., {\bf 63}, 840 (1989);
G.~L.~Kane, L.~T.~Wang and T.~T.~Wang,
Phys.\ Lett.\ B {\bf 536}, 263 (2002).

\bibitem{positrons2}
E.~A.~Baltz, J.~Edsjo, K.~Freese and P.~Gondolo,
Phys.\ Rev.\ D {\bf 65}, 063511 (2002).

\bibitem{posbaltz}
E.~A.~Baltz and J.~Edsjo,
Phys.\ Rev.\ D {\bf 59} (1999) 023511
[arXiv:astro-ph/9808243].

\bibitem{heat1995}
S.~W.~Barwick {\it et al.}  [HEAT Collaboration],
Astrophys.\ J.\  {\bf 482}, L191 (1997)
[arXiv:astro-ph/9703192].

\bibitem{heat2000}
S.~Coutu {\it et al.}  [HEAT-pbar Collaboration],
``Positron Measurements With the HEAT-pbar Instrument'',
in Proceedings of 27th ICRC (2001).



\bibitem{heat}
S.~Coutu {\it et al.},
Astropart.\ Phys.\ {\bf 11}, 429 (1999), 
[arXiv:astro-ph/9902162].



\bibitem{pythia}
T.~Sjostrand, P.~Eden, C.~Friberg, L.~Lonnblad, G.~Miu, S.~Mrenna and E.~Norrbin,
Comput.\ Phys.\ Commun.\  {\bf 135}, 238 (2001)
[arXiv:hep-ph/0010017].

\bibitem{darksusy}
P.~Gondolo, J.~Edsjo, L.~Bergstrom, P.~Ullio and E.~A.~Baltz,
arXiv:astro-ph/0012234;
http://www.physto.se/~edsjo/darksusy/.



\bibitem{diffusion}
W.~R.~Webber, M.~A.~Lee and M.~Gupta,
Astrophys.\ J.{\bf 390} (1992) 96.

\bibitem{lossrate}
M.~S.~Longair,
{\it High Energy Astrophysics}, Cambridge University Press, New York, 1994.

\bibitem{L}
D.~Maurin, F.~Donato, R.~Taillet and P.~Salati,
Astrophys.\ J.\  {\bf 555}, 585 (2001)
[arXiv:astro-ph/0101231];
D.~Maurin, R.~Taillet and F.~Donato,
Astron.\ Astrophys.\  {\bf 394}, 1039 (2002)
[arXiv:astro-ph/0206286].

\bibitem{2zonediffusion} 
F.~Donato, D.~Maurin, P.~Salati, R.~Taillet, A.~Barrau and G.~Boudoul,
Astrophys.\ J.\  {\bf 563}, 172 (2001);
D.~Maurin, R.~Taillet, F.~Donato, P.~Salati, A.~Barrau and G.~Boudoul,
arXiv:astro-ph/0212111.



\bibitem{eke}
V.~R.~Eke, J.~F.~Navarro, M.~Steinmetz, ApJ, 554, 114 (2001). 

\bibitem{hayashi}
E.~Hayashi, J.~F.~Navarro, J.~E.~Taylor, J.~Stadel and T.~Quinn,
ApJ, 584, 541 (2003).

\bibitem{islam} 
R.~R.~Islam, J.~E.~Taylor, and J.~Silk,
MNRAS, 340, 647 (2003). 

\bibitem{somerville} 
R.~S.~Somerville, T.~S.~Kolatt, MNRAS, 305, 1 (1999). 


\bibitem{tb2001} 
J.~E.~Taylor and A.~Babul, ApJ, 559, 716 (2001).

\bibitem{ts2003} 
J.~E.~Taylor and J.~Silk, MNRAS, 339, 505 (2003).

\bibitem{nomoore}
For example, Navarro {\it et al.}, submitted to MNRAS (2003), [arXiv:astro-ph/0311231].

\bibitem{tbprep}
J.~E.~Taylor and A.~Babul, MNRAS, (2003) in press,
[arXiv:astro-ph/0301612], and references therein.





\bibitem{secbg}
I.~V.~Moskalenko and A.~W.~Strong, 
Astrophys.~J.~{\bf 493}, 694 (1998).







\bibitem{nonthermal}
R.~Jeannerot, X.~Zhang and R.~H.~Brandenberger,
JHEP {\bf 9912}, 003 (1999)
[arXiv:hep-ph/9901357];
R.~Kallosh, L.~Kofman, A.~D.~Linde and A.~Van Proeyen,
Phys.\ Rev.\ D {\bf 61}, 103503 (2000)
[arXiv:hep-th/9907124];
G.~F.~Giudice, I.~Tkachev and A.~Riotto,
JHEP {\bf 9908}, 009 (1999)
[arXiv:hep-ph/9907510];
G.~F.~Giudice, A.~Riotto and I.~Tkachev,
JHEP {\bf 9911}, 036 (1999)
[arXiv:hep-ph/9911302];
T.~Moroi and L.~Randall,
Nucl.\ Phys.\ B {\bf 570}, 455 (2000)
[arXiv:hep-ph/9906527];
M.~Fujii and K.~Hamaguchi,
Phys.\ Lett.\ B {\bf 525}, 143 (2002)
[arXiv:hep-ph/0110072];
M.~Fujii and K.~Hamaguchi,
arXiv:hep-ph/0205044;
M.~Fujii and M.~Ibe,
arXiv:hep-ph/0308118.




\end{thebibliography}
\end{document}